\documentclass[useAMS,usenatbib,fleqn]{mnras}

\usepackage{txfonts}
\usepackage[pdftex]{graphicx}
\usepackage{times}
\usepackage[disable]{todonotes}

\newcommand{\hmsun}{h^{-1}{\rm M}_\odot}
\newcommand{\hmpc}{h^{-1}{\rm Mpc}}

\newcommand{\kms}{{\rm ~km~s^{-1}}}

\defcitealias{dgl_coherencia_2016}{GL16}

\title[A scenario for cosmic void motions]{The sparkling Universe: 
                                           a scenario for cosmic void motions}

\author[Ceccarelli et al.]{
\parbox[t]{\textwidth}
{Laura Ceccarelli, Andr\'es N. Ruiz, Marcelo Lares, Dante J. Paz, 
Victoria E. Maldonado, Heliana E. Luparello \& Diego Garcia Lambas}
\vspace*{6pt}\\
Instituto de Astronom\'{\i}a Te\'orica y Experimental (IATE), CONICET-UNC\\
and Observatorio Astron\'omico, Universidad Nacional de C\'ordoba, Argentina
}

\begin{document}

\date{\today}

\maketitle

\begin{abstract}
Cosmic voids are prominent features of the Universe, encoding relevant
information of the growth and evolution of structure through their dynamics.
Here, we perform a statistical study of the global motion of cosmic voids using
both a numerical simulation and observational data.  
Their relation to large-scale mass flows and the physical effects that drive
those motions.  We analyse the bulk motions of voids, finding void mean bulk
velocities in the range $300$ to $400\kms$, depending on void size and the
large-scale environment.  
Statistically, small voids move faster, and voids in relatively higher density
environments have higher bulk velocities. 
Also, we find large-scale overdensities (underdensities) along (opposite to)
the void motion direction, suggesting that void motions respond to a pull-push
mechanism.  
Our analysis suggests that their relative motions are generated by large-scale
density fluctuations. 
In agreement with linear theory, voids embedded in low (high) density regions
mutually recede (attract) each other, providing the general mechanism to
understand the bimodal behavior of void motions. 
We have also inferred void motions in the Sloan Digital Sky Survey using linear
theory, finding that their estimated motions are in qualitatively agreement
with the results of the simulation. 
Our results suggest a scenario of galaxies and galaxy systems flowing away from
void centers with the additional, and more relevant, contribution of the void
bulk motion to the total velocity.
\end{abstract}

\begin{keywords}
Cosmology: large scale structure of Universe -- Cosmology: observations -- 
Methods: data analysis -- Methods: statistics
\end{keywords}


\section{Introduction} 
\label{S_intro}

The large--scale flows in the Universe are directly related to the large mass
fluctuations  associated to the inhomogeneous galaxy distribution
\citep{Peebles}.  This
distribution is dominated by large virialized clusters connected by filaments
and unveils the presence of large--scale underdense regions widely known as
cosmic voids. 

This large--scale, underdense regions in the galaxy distribution are also
present in dark matter and halo distributions \citep[see, e.g., ][and
references therein]{bertschinger_self-similar_1985, sheth_hierarchy_2004,
hoyle_voids_2004, ceccarelli_voids_2006, aragon-calvo_unfolding_2010,
sutter_first_2012}.  Based on the fact that dark matter haloes trace the galaxy
distribution, \citet{padilla_spatial_2005} have studied and compared voids
defined by dark matter, haloes and galaxies, finding that these different
tracers produce voids with similar spatial and dynamical properties.  

On the other hand, the distributions of mass and voids are strongly connected
\citep{white_1979} and therefore void statistics can provide a simplified way
to extract information from the clustering pattern, giving clues on the
formation and evolution of the cosmic web. 
Thus, voids have an active interplay with large--scale flows affecting the
formation and evolution of structures in the Universe.  These large--scale
underdensities exhibit local expansion which in some cases, depending on the
large--scale environment \citep{sheth_hierarchy_2004,paz_cluesII_2013}, can be
reverted to collapse at larger scales, generating   global convergent or
divergent flows.

The peculiar velocity field can be directly related to the mass distribution
by applying linear theory \citep{Peebles}, a fact that can be used to
reconstruct it in the observations \citep{wang_reconstructing_2009}.
These large--scale flows can be understood as the result of the process of
gravitational instability with overdense (underdense) regions attracting
(repelling) material.  
In order to assess the bulk motion of voids we use linearized velocity field to
derive the core and shell void bulk motions as in \citet[hereafter
GL16]{dgl_coherencia_2016}.

Several works have focused on void properties, their dynamics and spatial
properties, but their possible bulk motions have received little attention.
Nevertheless, large--scale bulk flows have been analyzed using the peculiar
velocity field in the nearby Universe \citep{watkins_consistently_2009,
watkins_cosmicflows_2015}.

In \citetalias{dgl_coherencia_2016} we studied the pairwise velocity of voids
in simulations as well as in observational data finding that voids exhibit non
negligible coherent motions in the Universe.  
Here we present a complementary study of such global motions, focusing into the
details of the statistical properties of void velocities and the causes of the
void motions. 
This paper is organized as follows:
in Section \ref{S_data} we describe the data used which comprises the
observational and simulated void catalogues, and we include a brief description
of the void finder algorithm.
In Section \ref{S_mov_sim} we provide an analysis of void motions in
the simulation, where we examine the dependences of void velocity and the
effects of density around voids on their movement.  We also provide a more
detailed analysis of the coherence of the pairwise void velocities introduced
in \citetalias{dgl_coherencia_2016} in subsection \ref{ss_coherence}.
In section \ref{S_vel_SDSS} we analyze velocities of voids in SDSS and perform
a comparison of observational and theoretical results on void motions. 
Finally, we discuss our results in Section \ref{S_concl}.


\section{Void catalogues} \label{S_data}

\subsection{Data sources: galaxy catalogue and simulation}

The observational sample of galaxies used in this work was obtained form a
volume complete sample extracted from the Main Galaxy Sample
\citep{strauss_spectroscopic_2002} of the Sloan Digital Sky Survey Data Release
7 (SDSS-DR7).  This sample is defined by a limiting redshift $z=0.12$ and a
maximum absolute magnitude in the $r-$band $M_r=-20.3$. 
The parameters defining this galaxy sample are chosen by requiring to be
sufficiently dense to minimize shot noise and provide an accurate
identification, and large enough to contain a suitable number of voids to
obtain statistically significant results.
In order to derive void mean peculiar velocities we adopt the linearized
velocity field reconstruction by \citet{wang_reconstructing_2012}, who used
observed groups of galaxies as mass tracers within the framework of linear
theory and apply the linear relation between mass overdensity and peculiar
velocity \citep{wang_reconstructing_2009} to associate peculiar velocities to
the SDSS regions occupied by voids. 

On the numerical side, we use haloes from an $N$-body dark matter simulation
with $1024^3$ particles in a cubic volume of $1h^{-1}$Gpc on a side. The
cosmological parameters $\Omega_M=0.279$, $\Omega_\Lambda=0.721$,
$\Omega_b=0.0462$, $h=0.7$, $n=0.972$ and $\sigma_8=0.821$ correspond to a
$\Lambda$ Cold Dark Matter ($\Lambda$CDM) model in concordance with WMAP9
results \citep{hinshaw_wmpa9_2013}.
Initial conditions were generated using the public code \textsc{music}
\citep{hahn_music_2011}.  The evolution of the simulation until $z=0$ was
performed with the public version of \textsc{gadget-2}
\citep{springel_gadget2_2005}. 
The simulation was processed with \textsc{rockstar}
\citep{behroozi_rockstar_2013}, identifying $3983265$ dark matter haloes as
bound structures with at least $20$ particles.


\subsection{Void identification} \label{SS_voidcat}

The algorithm applied to identify voids is described in
\cite{ruiz_cluesIII_2015}, which is a modified version of the procedures
presented in \cite{padilla_spatial_2005} and \cite{ceccarelli_voids_2006}.
The void finding algorithm selects underdense regions as void candidates, the
larger spherical region centered in a void candidate position satisfying
$\Delta(R_{\rm void}) < -0.9$ is selected as an underdense sphere, being
$\Delta$ the integrated density contrast and $R_{\rm void}$ the void radius.
Of all underdense spheres selected, the larger ones which do not overlap with
any other are identified as voids. 

We apply our void finding algorithm to the haloes in the numerical simulation
and to the volume limited sample of galaxies. 
The final void catalogues contain $13430$ voids in the halo distribution and
$363$ galaxy voids.
We notice that \citet{wang_reconstructing_2012} conclude that the bias
in the reconstruction of the SDSS peculiar velocity field is small in
the inner region of the survey.  The volume for which this field is
reliably determined by this method is approximately two thirds of the
total volume of the main galaxy sample.  Consequently, we have considered 245
voids, identified in the observational data restricted to this smaller region.


\section{The motion of voids} \label{S_mov_sim}

Void internal dynamics is characterized by a bimodal behaviour which strongly
depends on the large--scale environment
\citep{sheth_hierarchy_2004,paz_cluesII_2013}. 
Although the central regions of all voids show similar dynamics consistent with
radial expansion, the surrounding mass distribution determines the velocity
field at void shells, making voids dynamically connected to the cosmic web
\citep{ruiz_cluesIII_2015}. 

It has been widely assumed that voids are at relative rest with respect to the
comoving coordinate system, however this topic has not been analyzed in detail.
The motions of voids as full entities have been noticed by
\citet{gottlober_structure_2003}, although on the other hand, it has been
suggested that voids may have negligible bulk velocities
\citep{sutter_life_2014}.  
We notice that it is difficult to compare possibly contradictory results on a
common basis due to the different type of void identification schemes involved.
Void bulk velocities are influenced by large-scale density contrasts, so that
voids identified with different underdensity thresholds could have different
velocities. 
Recently, \citet{wojtak_time_2016} have reported void motions using a
voidfinder analog to ZOBOV \citep{neyrinck_2008}. It is worth mentioning that
void motions are also detected using different methods.
In a recent paper, we have determined the existence of a bimodal behaviour of
voids motions \citepalias{dgl_coherencia_2016}.
Inspired on these results, in this work we analyze the dynamics of voids as
global entities considering the peculiar velocity field as well as their
connection with the large--scale mass distribution.
In this section we show that voids are non--static components of the
large--scale Universe, exhibiting significant peculiar motions. 


\subsection{The large-scale environment of voids and its relation to void global motions}  
\label{S_lssvel}

  \begin{figure*}
  \includegraphics[width=1.0\textwidth]{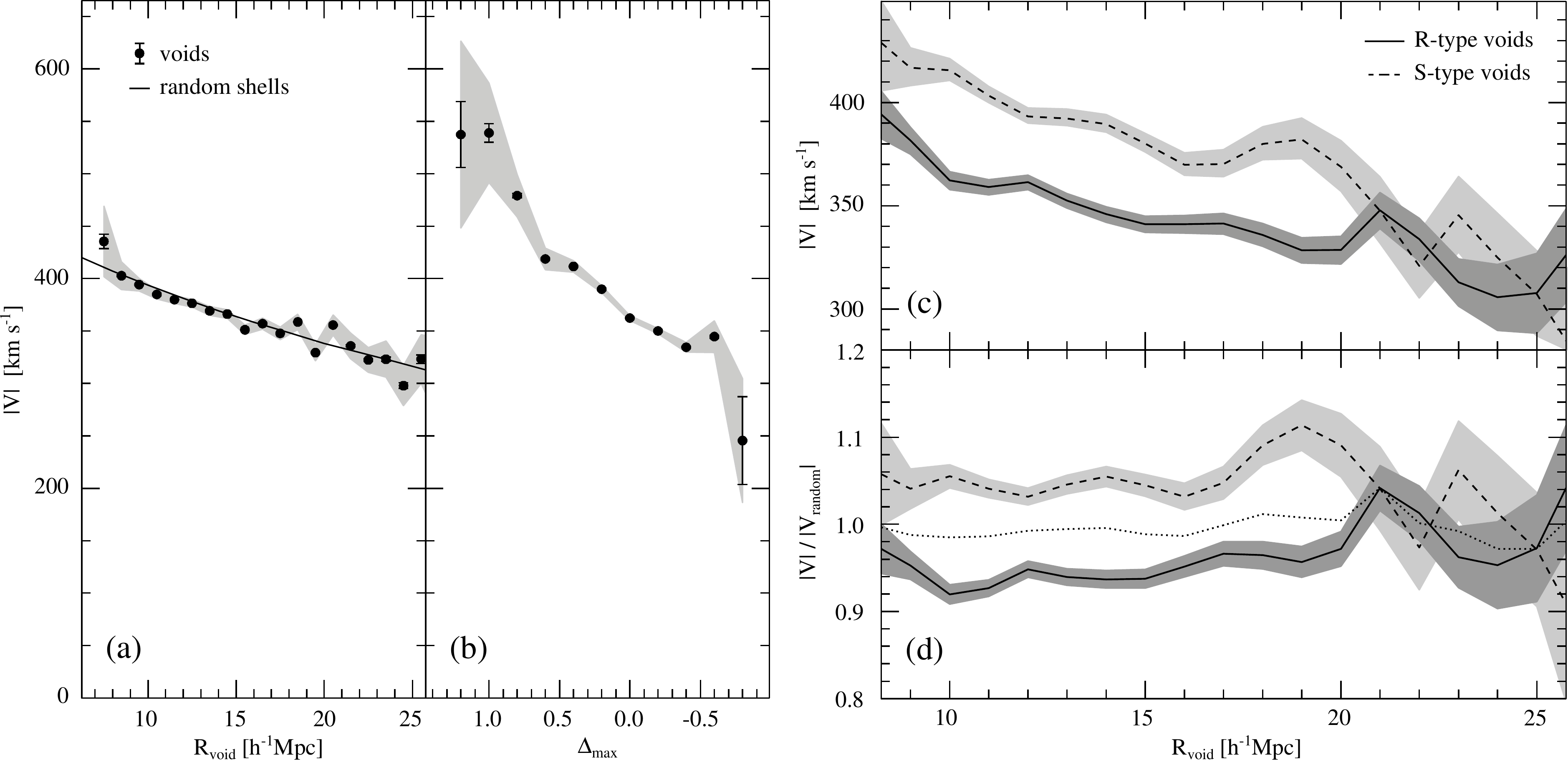}%
  \caption{
  {\it Left:} 
  In panel (a) we show the mean velocity for voids in the simulation box as a
  function of void radii (black points). The solid black line indicates the
  mean velocity of randomly centered spheres. Panel (b) shows the mean velocity
  of voids as a function of $\Delta_{\rm max}$ (black points).  In both panels
  the error bars indicate the standard error on the mean velocity and the grey
  regions shows the standard deviation of the velocity distribution.
  {\it Right:} In panel (c) we show the mean velocity as a function of the void
  radius for S-type (dashed line) and R-type voids (solid line) in the
  simulation. Panel (d) shows the ratio between the mean velocities of
  void and random spheres as a function of size for the full void sample
  (dotted line), S-type voids (dashed line) and R-type voids (solid line) in
  the simulation. The grey shaded regions represent the standard deviation of 
  the velocity distributions in both panels.
  }
  \label{fig:velm} 
  \end{figure*}
%

Throughout  this paper we use the void--shell mean velocity instead of that of
void cores. This is suitable since by definition there are very few galaxies
(haloes) inside voids and we have shown previously that shell bulk velocities 
trace well the void core motions \citepalias{dgl_coherencia_2016}. 
According to this, to compute the void velocities we calculate the mean bulk 
velocity of haloes located at void--centric distances between $0.8$ and $1.2$ 
void radius, which correspond to the denser shell surrounding voids.

Panel (a) of Fig. \ref{fig:velm} shows the mean void velocities as a function
of void radius in the simulation. 
Black circles show the complete sample of simulated voids, the error bars
represent the error of the mean and the region enclosed by the grey solid lines
indicates the velocity dispersion. 
It can be seen in this figure a clear trend of smaller velocities for large
voids, smaller ones ($R_{\rm void} < 10\hmpc$) exhibiting velocities as large
as $400\kms$, decreasing to $300\kms$ for the largest voids.
The black solid line in the figure indicates the mean velocity of randomly
placed shells in the simulation box. 
It can be noticed that the motions of voids and randomly placed shells are
similar, showing that voids are equally affected by the large--scale flows
originated by the large-scale structure.
Taking into account that the sources of peculiar velocities are mass density
fluctuations and the presence of large--scale mass correlations, we have
compared the motion of voids embedded in different global density environments.
For this purpose, we have analysed the dependence of void bulk velocities on
the surrounding large--scale density of haloes. 
In panel (b) of Fig. \ref{fig:velm} we show the void mean velocity as a
function of maximum overdensity at void environment ($\Delta_{\rm max}$) which,
for each void, is the absolute maximum of the integrated density profile at
distances between $2$ and $3$ $R_{\rm void}$ to the void center (see
\citet{ceccarelli_cluesI_2013} for a more detailed explanation of $\Delta_{\rm
max}$). 
As it can be seen in the figure, void mean velocities show a positive trend,
ranging approximately $300\kms$ for voids with $\Delta_{\rm max} \sim -0.5$ to
velocities higher than $500\kms$ for voids with $\Delta_{\rm max} > 1.0$.

In order to disentangle the effects of void environment and size on velocities,
we have also analyzed the velocity dependence on void radius for voids in low
and high density regions, namely R- and S-type voids, separately
\citep{ceccarelli_cluesI_2013,paz_cluesII_2013}. 
We notice that our void classification is actually based on the close void
environment, where R-type voids are surrounded by underdense large--scale
regions ($\Delta_{\rm max} < 0$) and S-type voids corresponds to voids embedded
in overdense regions ($\Delta_{\rm max} > 0$). This classification is closely
related to void evolution and describes in a simple way the two different
modes, namely the void--in--void and void--in--cloud processes
\citep{sheth_hierarchy_2004}.  
Motivated by these classical results, we have analyzed void evolution adopting
a criterion based on the density around voids, to select voids in overdense and
underdense regions (S-type and R-type voids respectively) in order to
associate them with the void--in--void and void--in--cloud pictures.

\begin{figure}
  \includegraphics[width=0.47\textwidth]{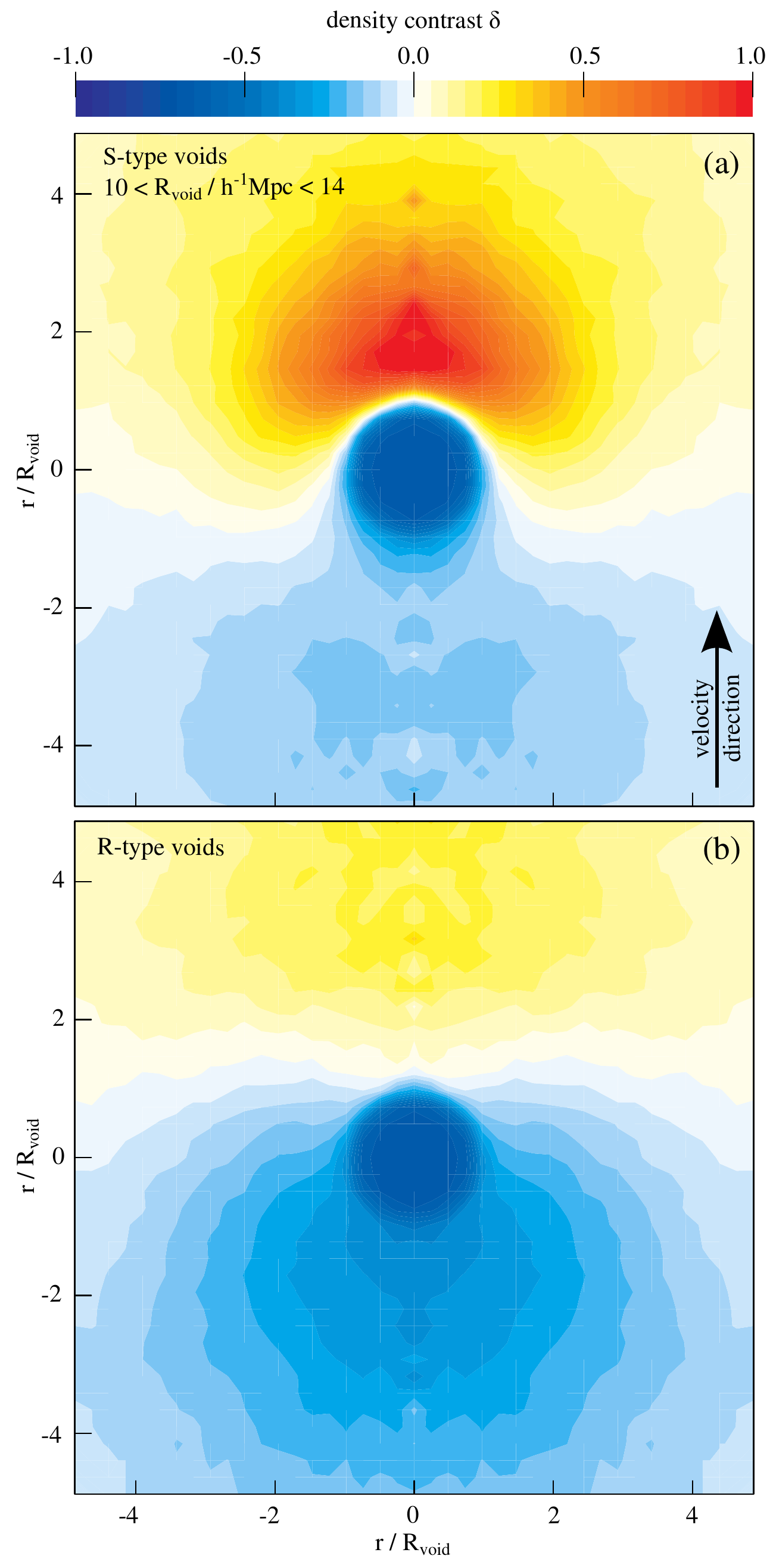}%
  \caption{
  Normalized stacked density contrast ($\delta$) around S-type (panel (a)) and
  R-type (panel (b)) voids in simulation for sizes in the range $10 < R_{\rm
  void} / \hmpc < 14$. Distances are normalized to void sizes and $y$-axis is 
  aligned to void velocity direction. Colorbar in the top shows the value of 
  the normalized stacked $\delta$.
  }
  \label{fig:stacking_voids} 
  \end{figure}

Given the large differences of the dynamics of shells surrounding voids
embedded in low and high density regions \citep{ruiz_cluesIII_2015} we can also
envisage to find differences of void bulk velocities of the two void types.
The mean bulk velocities are shown in panel (c) of Fig. \ref{fig:velm}, where
the dashed lines indicate the mean for S-type voids and the solid lines
correspond to the mean for R-type voids, the regions between grey lines
indicate the 1$\sigma$ deviations of the mean. 
We show the results for void radii in the range 10-25 $\hmpc$, suitable to
compare the velocity of both types of voids with statistically significant
number of voids. 
We notice that we obtain systematic larger mean velocities for S-type voids and
reach the highest values for small S-type voids, $\sim  400\kms$ for $R_{\rm
void}< 12\hmpc$. Large R-type voids have mean bulk velocities below $350\kms$.
Besides the statistical dependence of void size on the surrounding density, the
magnitude of mean void velocity is related to both, void size and environment. 
This behaviour could be a natural outcome of the stronger gravitational pull of
the dense regions surrounding S-type voids (as it will be seen in the
discussion of Fig. \ref{fig:stacking_voids} bellow).

We have also compared the mean bulk velocity of voids in different large--scale
environments with that of randomly placed spheres. 
For both voids and random spheres, the bulk velocity is obtained by calculating
the mean velocity of dark matter haloes in shells of radius $R_{\rm void}$.
The ratio of void/random sphere velocities as a function of void size are
displayed in panel (d) of Fig. \ref{fig:velm}, for R-type voids (solid line),
S-type voids (dashed line) and the full sample of voids (dotted line). 
As it is can be appreciated in this figure, randomly placed spheres velocities
show an intermediate behaviour between S-type and R-type voids.  
Since the random spheres sample should comprise a mixture of overdense and
underdense regions, it is expected an intermediate behaviour between the
overdense and underdense void shells.

  \begin{figure}
  \includegraphics[width=0.47\textwidth]{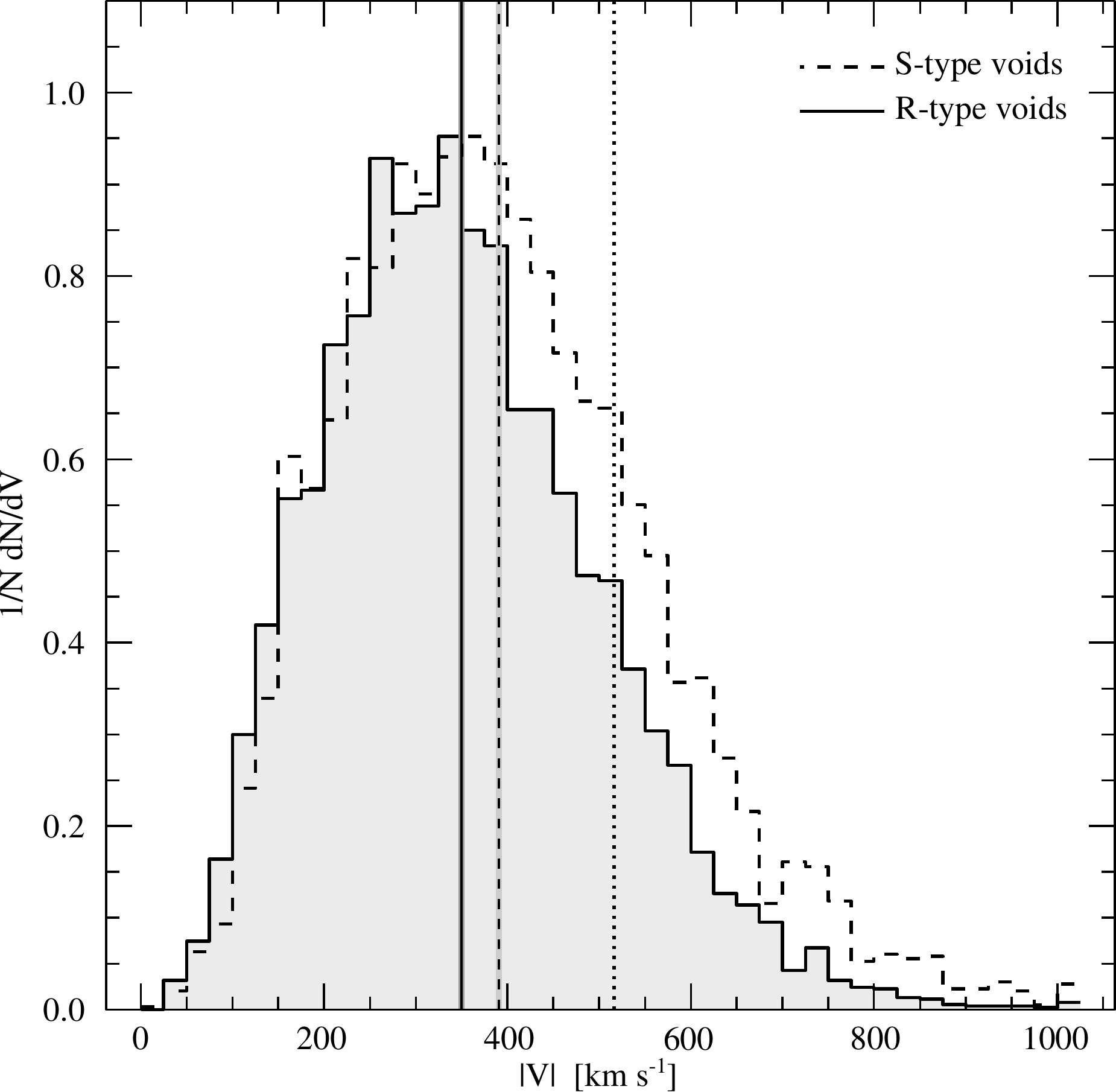}%
  \caption{
  Void velocity normalized distributions for the voids identified in the
  numerical simulation. S-type voids are represented with the dashed line and
  R-type voids with the solid line. Vertical solid (dashed) line shows the mean
  velocity value for R-type (S-type) voids. Shaded bands around these lines 
  represent the standard error of the mean values. Vertical dotted line indicates
  the mean velocity of haloes having $M > 10^{12}\hmsun$.
  }
  \label{fig:hist_simu} 
  \end{figure}

In Fig. \ref{fig:hist_simu} we display the magnitude of the void velocity
distributions for S- and R-type voids where it can be seen that their
velocities span a wide range of values, ranging from $0$ to $1000\kms$. By a
more detailed inspection, it can be noticed that S-type voids tend to have
velocities systematically larger than R-type voids, with an excess of voids
with velocities larger than $800\kms$, and up to $1000\kms$ for S-type voids
(dashed line), whereas R-type voids remain mostly below $800\kms$ (solid line).
Vertical lines in Fig. \ref{fig:hist_simu} show the mean void velocity for
S-types (dashed) and R-types (solid), corresponding to $390\kms$ and $350\kms$
respectively. Grey shaded bands represent the error of the mean. 
The dotted vertical line at $515\kms$ show the mean velocity of dark matter 
haloes. 
We have selected the haloes having $M>10^{12} \hmsun$ in the simulation box
and we calculate their averaged velocity obtaining a mean of $515 \kms$, which is
consistent with those obtained by \citet{pivato_infall_2006}. 
It is remarkable that mean void and halo velocities are of the same order
despite their very different nature, haloes being the most compact, extremely
dense objects, and voids the largest empty regions in the Universe.

In order to shed light on the sources of the motion of voids we have explored
the relation between the direction of bulk motion and the mass density
distribution surrounding voids.  For this aim, we have stacked the mass density
distribution surrounding voids aligning with the velocity vector and
normalizing the distances to void radii.
The resulting stacking has axial symmetry with respect to the void velocity
vector and so we average the signal around this direction.
In Fig. \ref{fig:stacking_voids} we show the normalized density maps of
stacked voids where the $y$-axis direction correspond to that of voids bulk
motion. 
In panel (a) we display the results for S-type voids, and in panel (b) for
R-type voids.  Both subsamples have $10 < R_{\rm void} / \hmpc < 14$, a range
selected in order to have enough number of R- and S-type to enhance
visualization, although the same pattern occurs for different size ranges.
The normalized density contrast ($\delta$) increases from blue to red and white
color corresponds to the mean density, as it is indicated in the color bar.
It can be seen that for both void types the motion points towards
overdensities, being larger for the case of S-type voids. 
This is expected due to the definition of S-type voids, which are embedded in
regions with higher densities ($\Delta_{\rm max}>0$).
We also notice that void are pushed away from underdense regions as expected if
voids follow the global flow pattern driven by large--scale mass fluctuations. 
In this case, R-type voids present a more noticeable underdensity opposite to
the direction of void motions.
The general picture of void motions can be understood in terms of a pull and
push mechanism driven by the large--scale velocity flows, where S-type voids in
high density environments are pulled by the collapsing structures and R-type
voids are pushed away from the expanding underdense environment in which they
are embedded.


\subsection{Coherence of pairwise large-scale void motions}\label{ss_coherence} 

  \begin{figure}
  \includegraphics[width=0.47\textwidth]{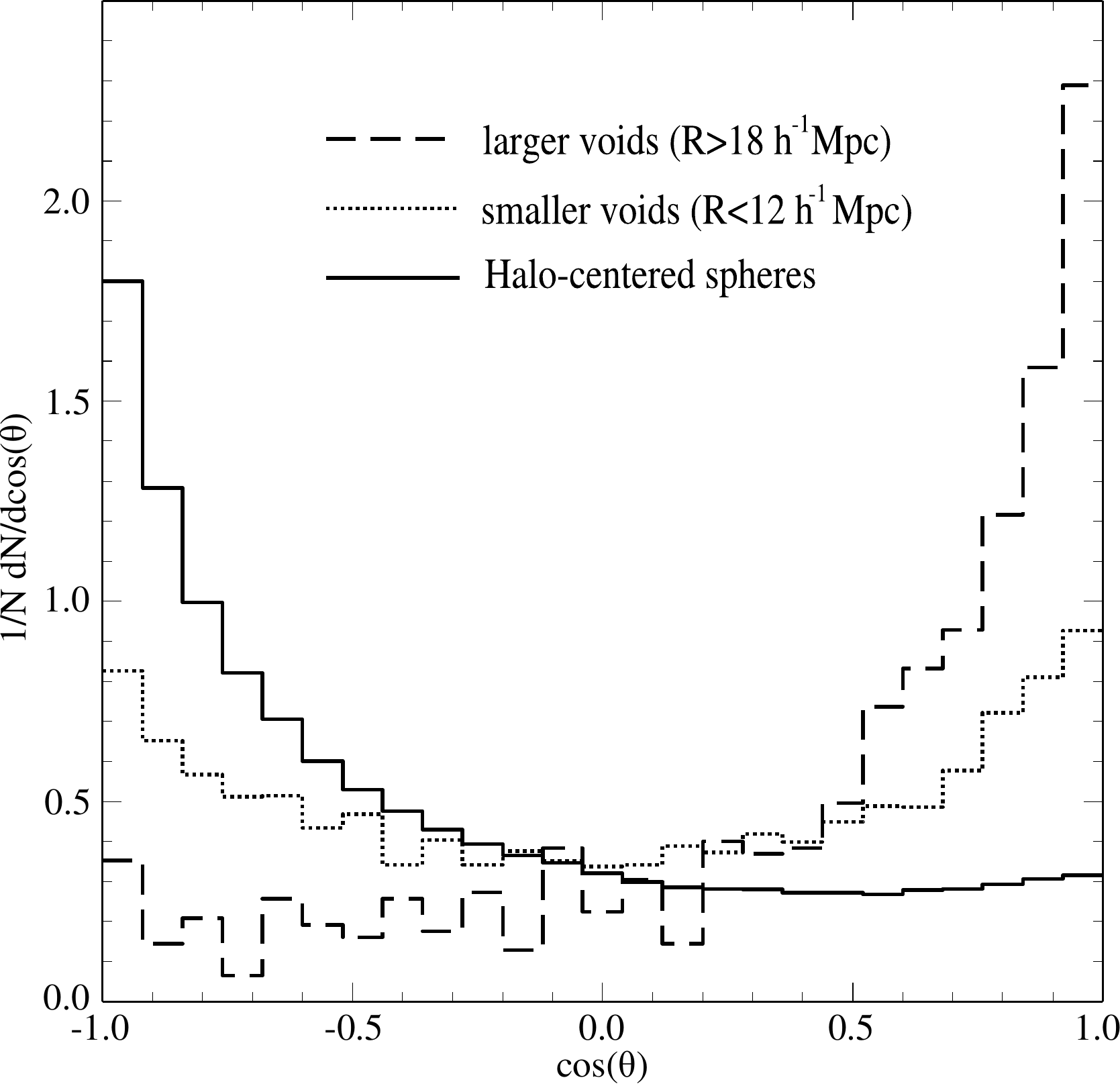}%
  \caption{
  Number counts of pairs of spheres having velocities with relative
  orientations given by $\cos(\theta)$.
  The dashed line corresponds to large voids in the void catalogue ($R_{\rm
  void} > 18 \hmpc$), the dotted line corresponds to the small voids ($R_{\rm
  void} < 12 \hmpc$) and the solid line corresponds to spheres around high mass
  haloes ($M > 10^{14} \hmsun$), with the same size distribution than
  that of the void sample.
  }
  \label{fig:cos} \end{figure}

In \citetalias{dgl_coherencia_2016} we reported two void populations according
to their large--scale environment: voids mutually receding and those mutually
approaching. 
In this work, it was also analysed the connection between the
pairwise motion bimodality and void classification, finding S-type void pairs
systematically approaching each other, and R-type voids mutually receding.

Upcoming surveys such as Hobby-Eberly Telescope Dark Energy Experiment
\citep[HETDEX,][]{hill_hetdex_2008}, Euclid \citep{euclid_2013} or Dark Energy
Spectroscopic Instrument \citep[DESI,][]{desi_2015} can lead to a revolutionary
progress on observational cosmology, since they will cover unprecedented large
scales, thus providing more significant samples of large voids.
In previous works \citep[see, for instance,][]{ceccarelli_cluesI_2013}, we
notice that the largest voids which are those most likely to be involved in
several cosmological tests, are of R-type and therefore are expected to have
systematically mutually receding motions, contrasting to the bimodal behaviour
of a mixed population of void types \citepalias{dgl_coherencia_2016}. 
It is expected that this behavior could be seen through the
distributions of $\cos(\theta)$ values, being $\theta$ the angle between the
difference velocity vector and the vector separation between voids as shown in
Fig. \ref{fig:cos}, where dashed lines and dot-dashed lines
correspond to large and small voids, respectively.
Note that in this scheme positive (negative) values of $\cos{\theta}$
indicate mutually receding (approaching) voids.  Then, large voids,
which are likely R-type, tend to receed and smaller voids, which are a
mixture of R and S-types, show a bimodal distribution of approching and
receeding voids.
For a suitable comparison, we also analysed regions centered in the most
massive haloes ($M > 10^{14} \hmsun$), considering that these systems will tend
to be embedded in large overdense environments.  As it is shown in solid lines
in Fig. \ref{fig:cos}, shells centered in massive haloes corresponding to
clusters of galaxies are likely to be approaching each other, similarly to
S-type voids, largely contrasting with the opposite behaviour of large, R-type
voids mutually receding.
%
  \begin{figure}  
  \includegraphics[width=0.47\textwidth]{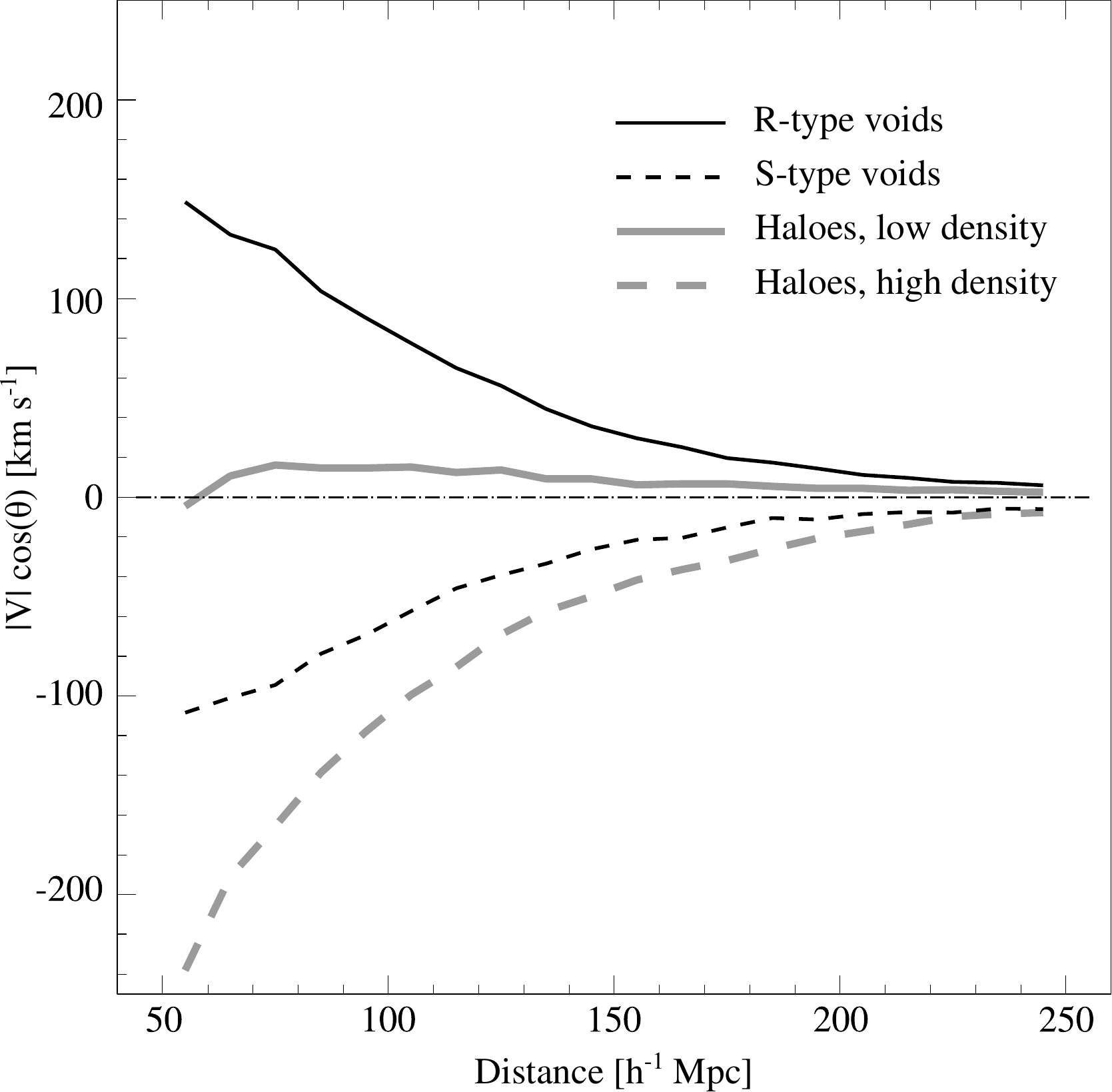}%
  \caption{
  Pairwise velocity ($|V|\cos(\theta)$) of spheres of radius $R$ as a function
  of the relative separation in the simulation.
  The solid and dashed black lines correspond to R-type and S-type void pairs,
  respectively.  
  The solid (dashed) grey lines correspond to spheres centered on haloes with
  $M > 10^{14} \hmpc$, with a total integrated density up to $3R$ lower
  (greater) than the mean, following the similar procedure as in voids.
  The sphere radius $R$ was taken at random from the same distribution of void
  radii.
  }
  \label{fig:vcos} \end{figure}

In order to perform a more detailed analysis of the relative motions of voids
and to put them within the large--scale context, we have compared the relative
pairwise velocity ($|V|\cos(\theta)$) as a function of relative separation for
spheres of radius $R$ centered in $M > 10^{14} \hmsun$ haloes, constrained to
reside in different density environments.
The results are shown in Fig. \ref{fig:vcos} for spheres classified into low
and high large--scale density, according to the integrated mass density up to 3
sphere radius similarly to the criterion applied to void classification into S-
and R-types.
It is worth mentioning that 97 per cent of these haloes are outside $1.2 R_{\rm
void}$ from any void so that the sample of massive halo--centered spheres trace
different regions from that of voids and allows us to analyse void motions in
the general context of global flows.
As a distinction to voids, which exhibit a bimodal behaviour of approaching and
recedding motions depending on the large-scale overdensity (Fig.
\ref{fig:vcos}, solid lines), spheres centered in the most massive haloes have
mainly mutually approaching motions. Also, independently of the constraints in
the global surrounding mass density, they do not show mean recedding motions,
as shown with dashed lines in Fig. \ref{fig:vcos}.

We also acknowledge that the filamentary structure surrounding voids is mostly
populated by relatively low mass haloes \citep{cautun_nexus_2013}, so their
global motion is related to that of the void core
\citepalias{dgl_coherencia_2016}.  
We have explored this behavior by studying the correlated pairwise velocities
for low and high halo mass ranges, defined by haloes having $M < 10^{14}\hmsun$
and $M < 10^{14}\hmsun$, respectively. We select haloes in low and high
large-scale environment, finding that the low mass halos exhibit a similar
behaviour than voids. Specifically, low mass halos in low density environments
exhibit receeding motions, similar to a pair of R-voids, whereas low mass halos
in high density regions attract each other, like S-type voids.

  \begin{figure}
  \includegraphics[width=0.47\textwidth]{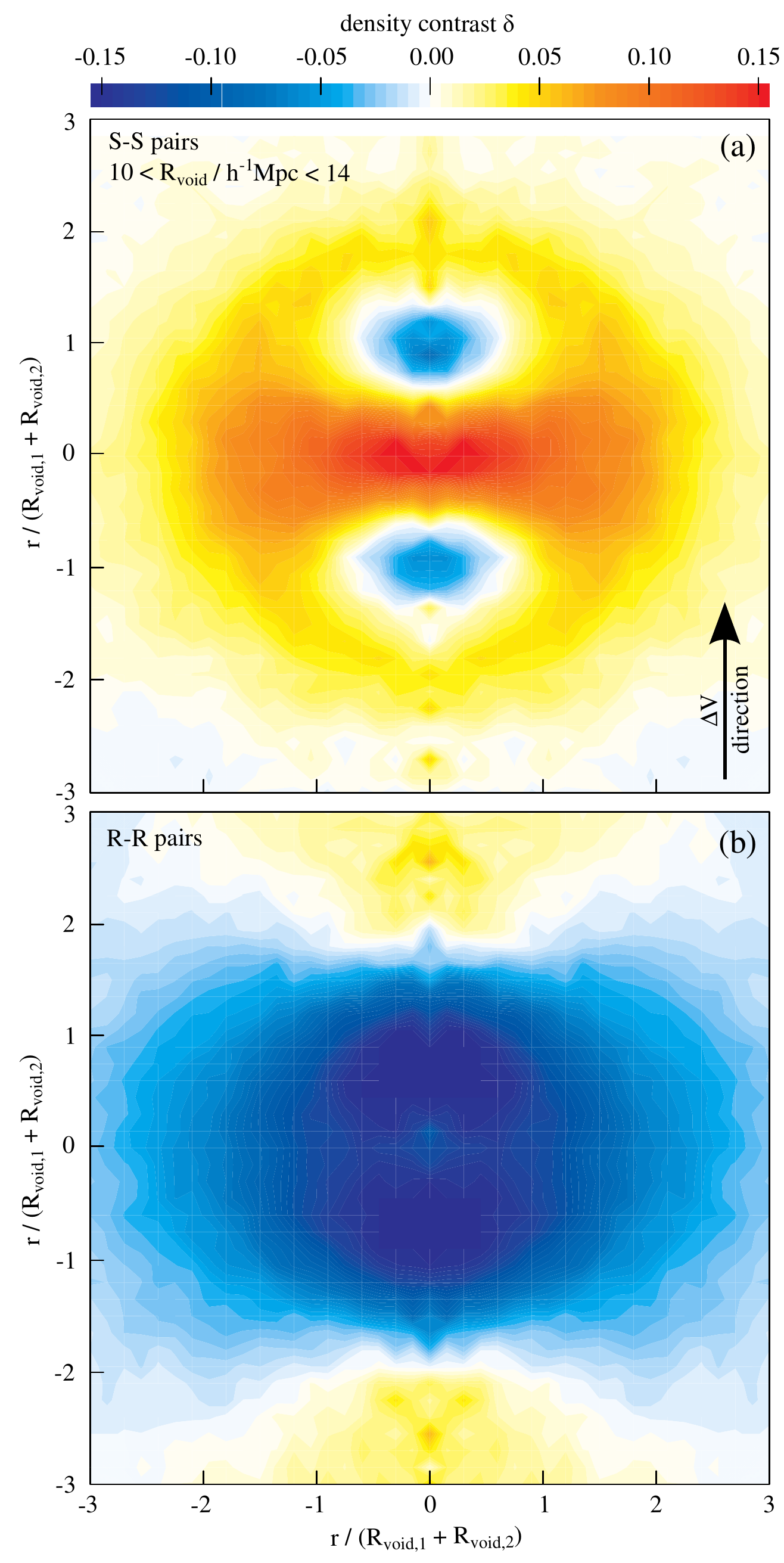}%
  \caption{
  Normalized staked density contrast ($\delta$) for S-S (panel (a)) and R-R
  void pairs (panel (b)). 
  Void pairs are selected from a subsample with sizes in the range $10 < R_{\rm
  void} / \hmpc < 14$ and a separation between 1.5 and 3 times the sum of their
  radii.
  Distances are normalized to the sum of void radii and the $y$-axis is
  oriented to the velocity difference direction. 
  As this direction is aligned with the relative separation direction, the
  coherent patter emerges. 
  Color bar in the top shows the value of the normalized stacked $\delta$.
  }
  \label{fig:stacking_pairs} \end{figure}

In Fig. \ref{fig:stacking_pairs} we show the halo density around void pairs
by stacking the pairs aligned to the vector of their relative velocity
difference. 
Panel (a) of Fig. \ref{fig:stacking_pairs} correspond to S-S void pairs,
meanwhile R-R pairs are shown in panel (b). In both cases, voids have a
separation distance in the range $1.5-3$ on units of the sum of the void radii
and with sizes from $10\hmpc$ to $14\hmpc$. 
As in Fig. \ref{fig:stacking_voids}, colours indicate the mass overdensity
values of the resulting stacking.
It can be noticed that the relative velocity of voids tend to be
aligned with the relative separation vector \citepalias{dgl_coherencia_2016},
for this reason the two bluest regions correspond to the largest underdensities
where voids are more likely located.

As it can be seen in the upper panel of Fig. \ref{fig:stacking_pairs}, there is
a noticeable overdensity between S-type void pairs. 
Consistent with mass fluctuation governing the peculiar velocity field, this
figure shows how the approaching motion of S-type void pairs found in
\citetalias{dgl_coherencia_2016} is driven by this overdense region residing
typically along the direction of their relative separation. 
By contrast, the density map of the lower panel of Fig.
\ref{fig:stacking_pairs} exhibit an underdense region between voids, and
overdensities in the opposite direction. 
This is consistent with a picture where voids are strongly affected by the
global environment, a larger fraction of small voids are likely to be in
overdense regions so that two close S-type voids will tend to approach each
other due to the action of the overdensity in which they are embedded. 
On the other hand, a pair of R-type voids, likely residing in a global
underdense region, will tend to be moving away from each other, pushed by the
underdense region between them, and pulled by denser regions at large
distances. 
These results provide further explanation to the coherence of void motions
presented in \citetalias{dgl_coherencia_2016}.

   

\section{The motion of observational voids} 
\label{S_vel_SDSS}

\begin{figure}
  \includegraphics[width=0.45\textwidth]{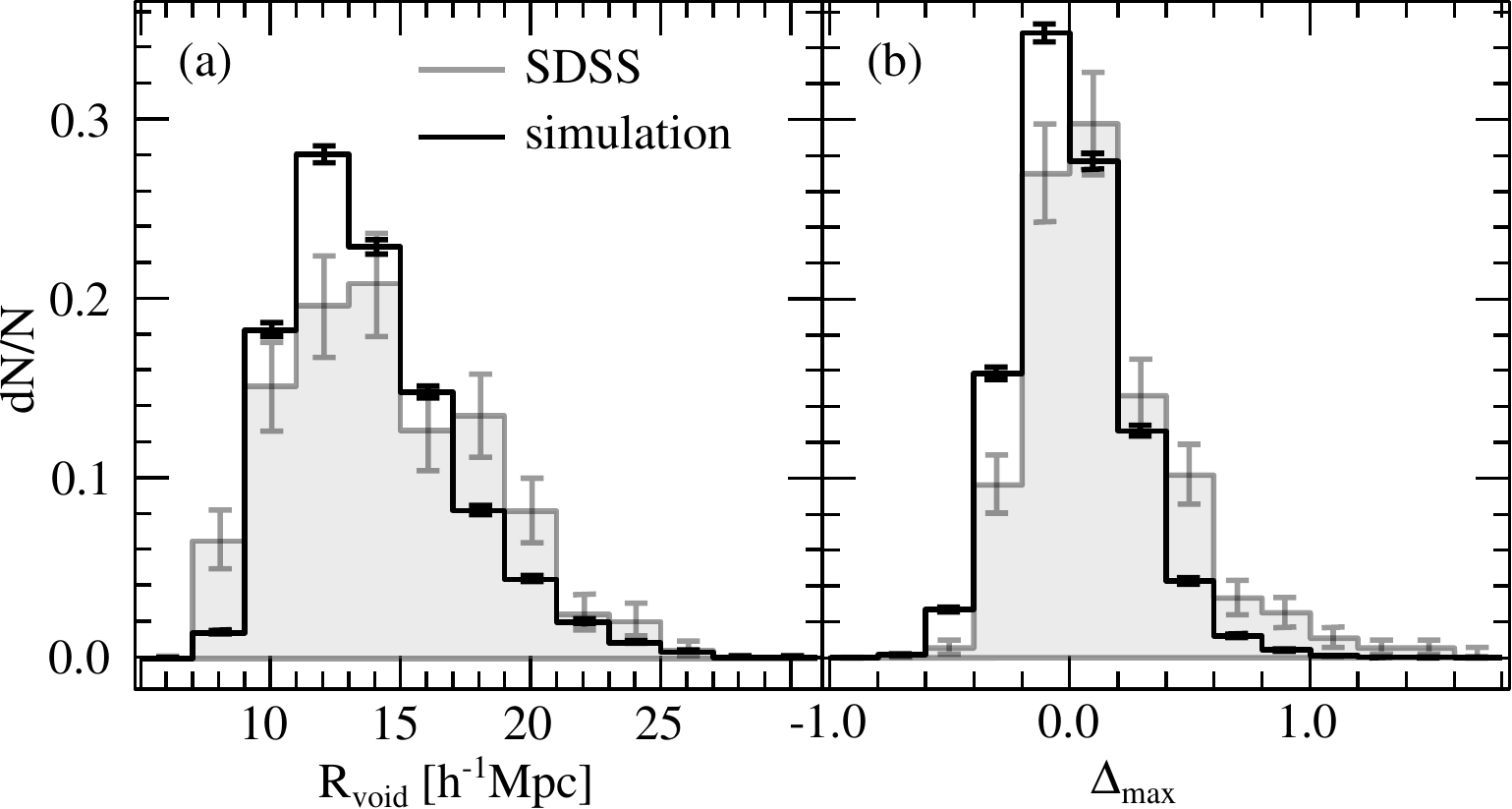}
  \caption{
   Normalized number counts of voids as a function of void radius
   ($R_{\rm void}$) in panel (a) and as a function of maximun
   overdensity ($\Delta_{\rm max}$) in panel (b).  In both panels voids
   identified in the SDSS-DR7 are in grey lines and voids identified
   in the numerical simulation in black lines. The errorbars for $dN/N$ 
   correspond to poisson uncertainties.
   }
  \label{fig:sdss_vs_simu} 
  \end{figure}

Once we have analysed into some detail void motion in the numerical simulation,
in this section our aim is to study the motions of observational voids in a
similar fashion.
To validate a direct comparison between the results obtained in the numerical
simulation with those we will present in the following sections for the
SDSS-DR7, in Fig. \ref{fig:sdss_vs_simu} we compare the normalized number
counts of voids as a function of void radius (panel (a)) and as a function the
maximum overdensity $\Delta_{\rm max}$ (panel (b)).
The distributions of voids identified in the SDSS-DR7 and in the numerical
simulation are plotted in grey and black lines, respectively. The
errorbars for $dN/N$ correspond to poisson uncertainties.
Due to the fact that both populations shares similar distributions of sizes and
$\Delta_{\rm max}$, we can expect similar properties of bulk motions in the
observational voids than those obtained previously in the numerical simulation.
 
\subsection{Void velocities} 

In order to assign peculiar velocities to galaxies in the observational
sample we adopt the linearized velocity field derived for the SDSS-DR7 region
by \citet{wang_reconstructing_2012}.
Velocities for observational voids are calculated following the same procedure
than in the simulation. In \citetalias{dgl_coherencia_2016} we have shown that
velocities of void shells in the numerical simulation suitably trace void
global motions. 
The validation of these results for observational data, and a justification of
the use of a linearized velocity field are presented in Appendix \ref{appA}.

  \begin{figure}
  \includegraphics[width=0.47\textwidth]{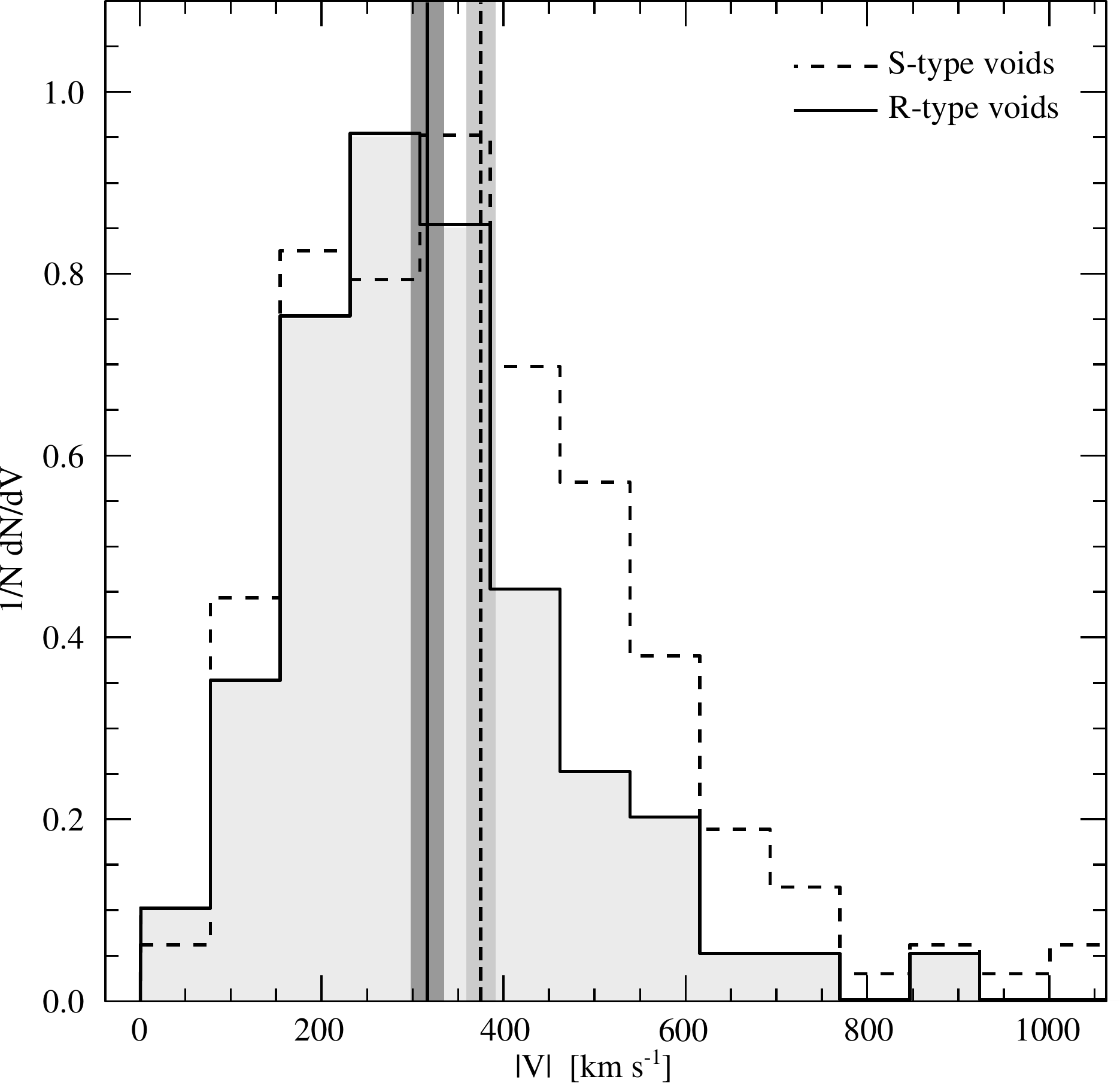}
  \caption{
  Void velocity normalized distributions for the voids identified in the
  SDSS data. S-type voids are represented with the dashed line and
  R-type voids with the solid line. Vertical solid (dashed) line shows the mean
  velocity value for R-type (S-type) voids. Shaded bands around these lines 
  represent the standard error of the mean values.
  }
  \label{fig:hist_sdss} 
  \end{figure}

In Fig. \ref{fig:hist_sdss} we display the void velocity distribution for S-
and R-type voids in SDSS. 
As it can be seen, the derived void bulk velocity distribution is similar to
that corresponding to voids in the simulation (see Fig. \ref{fig:hist_simu}),
ranging from $0$ to $1000\kms$ for R- and S-type voids.  
The same trend observed in the simulation is found here, where  S-type voids
are more likely to have larger bulk velocities than R-type voids, with an
excess of voids with velocities larger than $800\kms$ for S-type voids (dashed
line), whereas R-type voids remain mostly below $600\kms$ (solid line).
Vertical lines in Fig. \ref{fig:hist_sdss} indicate the mean void velocity
for S-type (grey dashed vertical line) and R-type (black dashed vertical
line), corresponding to $370\kms$ and $310\kms$, respectively.
These results are in qualitative agreement with those obtained in numerical
simulations (see Fig. \ref{fig:hist_simu}).

  \begin{figure}
  \includegraphics[width=0.47\textwidth]{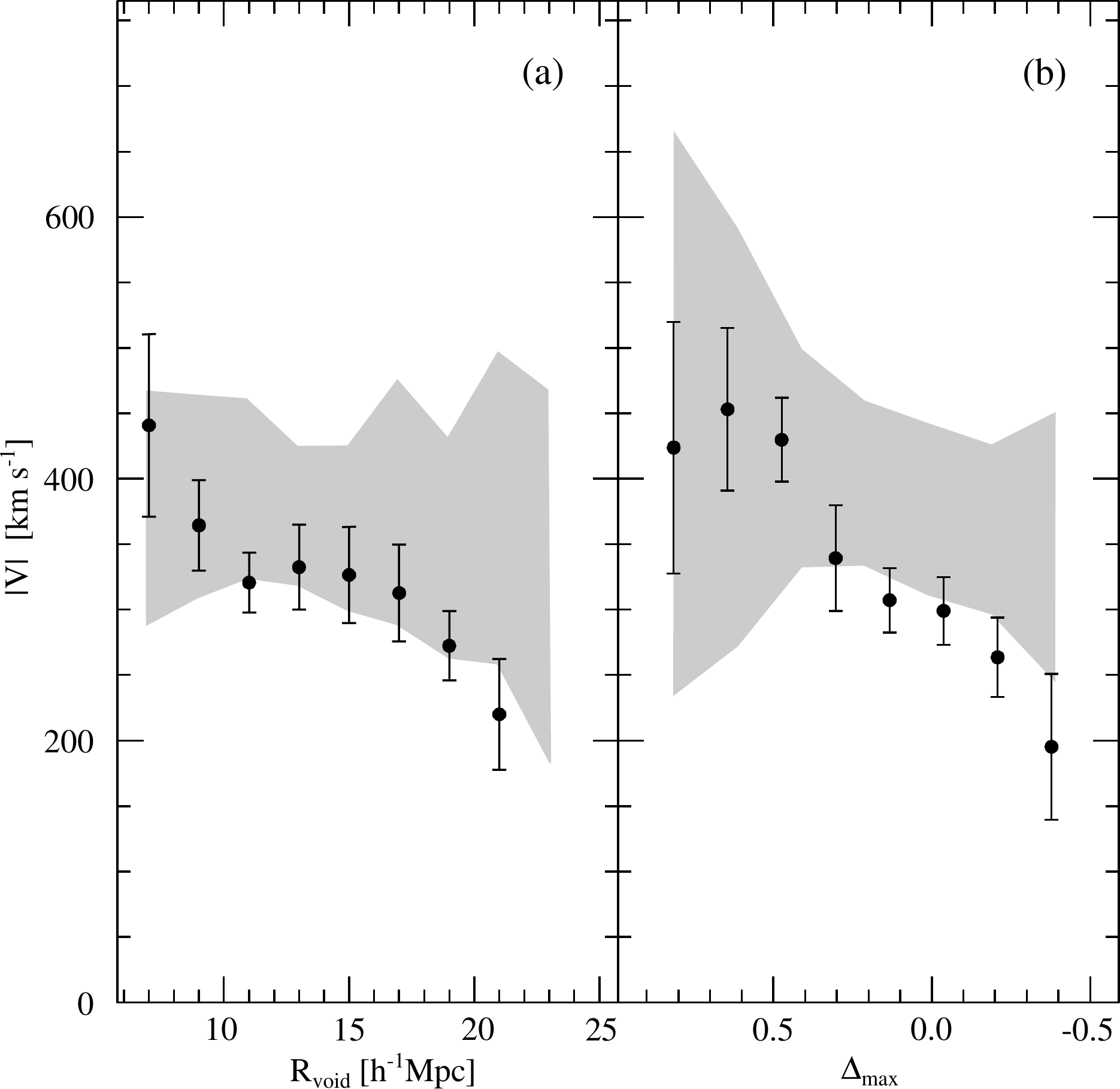}
  \caption{
  In panel (a) we show the mean bulk velocity of voids in the SDSS as a
  function of void radii and in panel (b) the same  mean bulk velocity as a
  function of $\Delta_{\rm max}$ (black points). In both cases, the error-bars
  correspond to the standard error of the mean and the grey shaded
  areas represent an estimation of cosmic variance (see text for details).   
  }
  \label{fig:velmsloan} 
  \end{figure}
%

We have also analysed the dependence of void velocity on void size and
$\Delta_{\rm max}$.
In panel (a) of Fig. \ref{fig:velmsloan} we show the mean void velocities as a
function of void radius in the SSDS. 
Black circles show the complete sample of SDSS voids, with error bars
representing the error of the mean. 
The grey shaded region represents an estimation of the cosmic variance
effect on the mean void velocities from simulations. We split the simulation
box into 64 independent realizations in the simulation with the same volume
than that spanned by the SDSS sample.
As it can be noticed in this panel, these trends are similar to those obtained
in the simulation where velocities become smaller as void size increases, from
$\sim500\kms$ to $\sim200\kms$ for voids with $7 < R_{\rm void}/ \hmpc < 21$.
In panel (b) of Fig. \ref{fig:velmsloan} we show the void mean velocity as a
function of maximum overdensity at void environment ($\Delta_{\rm max}$).
In this panel it can be seen that the mean velocities show a significant trend,
from approximately $200\kms$ for voids with $\Delta_{\rm max} \sim -0.5$ to
velocities higher than $400\kms$ for voids with $\Delta_{\rm max} > 1.0$.
It can be noticed that mean void velocities in the observations are in good
agreement with those predicted by simulations (grey shaded region).
We infer that voids in SDSS have significant bulk motions and have dependences
on void size and large--scale environment consistent with those found in the
simulation.
It is important to stress the suitable agreement between theoretical
and observational results, in particular, given the limitations of the
observational samples.

It is worth to mention that the possibility of using linearized velocity fields
to derive void bulk flows allows us to apply the method developed by
\citet{wang_reconstructing_2009} to diverse observational samples. 
This could be a powerful tool to analyse the large--scale flows traced by the
void velocities.   
This could be an interesting opportunity to carry out statistical studies of
the dynamic of voids in available galaxy surveys, without limiting  to the
nearby regions where peculiar velocities can be measured accurately.


\section{Discussion and conclusions}  \label{S_concl}

In this paper we explore into detail the phenomenology of void motions as
presented in \citetalias{dgl_coherencia_2016}. Voids exhibit significant bulk
motions and, in general, void and halo velocities follow the large flows
induced by large--scale mass fluctuations in the Universe. 
The interpretation of bulk flows is limited by the effects of small scale
motions \citep{watkins_cosmicflows_2015}. 
The advantage of analysing the large--scale bulk motions through void motions
is due to the simple internal dynamics that allow as to disentangle the local
and global velocity fields. 
In contrast to the case of spheres of the same volume centered in clusters or
randomly, void dynamics is governed by a divergent velocity field so that
galaxies (or haloes) flowing away from voids are subject to both void shell
expansion and bulk motion, indicating that void shells are in relative
expansion with respect to the central core region. 
Moreover, the radial outflow is not strongly turbulent, with comparable albeit
smaller, velocity dispersion in the shell surface \citep{padilla_spatial_2005}
with respect to the shell radial velocity.  The large bulk velocities of voids
in cosmological volumes found in this work imprint a global motion to those of
haloes in the shells.  
In this simple scenario, bulk flows can be thought of as a combination of void
(shell) global motions plus radial outflow and dispersion.
We notice that void (shell) bulk motions are of comparable magnitude to that of
haloes, in particular for those with $M<10^{14} \hmsun$). 
This is not totally surprising since void shells are populated by relatively
low-mass haloes so that the bulk velocities of the void shells are a major
contribution to halo motions superimposed to the void shell expansion.

Similarly, void pairwise velocities are consistent with that of large--scale
systems with motions generated by large--scale density fluctuations.  In
general, low (high) density regions mutually recede (attract) each other
consistent with a basic statement of Newtonian dynamics in an expanding world
model.  
Regions centered in high density peaks show approaching velocities in a similar
way than voids embedded in overdense regions.  On the contrary, very large, low
density regions, are populated by voids in the galaxy/halo distribution, and
are characterized by a global expansion. 
Our results on void pairwise motions, introduced on
\citetalias{dgl_coherencia_2016} and expanded here, provide an alternative
description of the evolution of the large-scale structure of the Universe. 
On this scenario, the coherent and non negligible bulk motions of voids are
supposed to play a fundamental role in the formation of structures
\citep{Boss_2014}.  Voids are affected by relative velocities of $\sim 400\kms$
at very large scales (up to 200$\hmpc$), these large scale flows should be
correlated with significant density fluctuations in regions larger than
100$\hmpc$ \citep{watkins_consistently_2009},  which are compatible with the
anisotropies on dark matter distribution around voids showed here (as it can be
seen in Figs. \ref{fig:stacking_voids} and \ref{fig:stacking_pairs}).
Such large scale anisotropies in matter distribution could be controversial
with predictions derived from Cosmic Microwave Background probes
\citep{Kashlinsky_2008,watkins_consistently_2009,Lavaux_flows_2010,
Feldman_flow_2010,Colin_2011,Ma_2015}.

Then, having knowledge of the coherence and non-negligible motions of voids is
crucial for observational cosmology.
For instance, the motion of galaxies and clusters at large scales
\citep{Nusser_flows_2011,Turnbull_flows_2012,Ma_Pan_2014,Hoffman_cosmic_bulk_2015}
could be governed by the large--scale streaming flows driven by cosmological
voids, generating significant bulk velocities at scales around a few hundred
$\hmpc$.  These motions, and the fluctuations associated to them, are not fully
compatible with predictions based on $\Lambda$CDM cosmological models.

We also forecast the impact of these results on cosmological tests using voids
in forthcoming large-scale surveys such as HETDEX \citep{hill_hetdex_2008},
Euclid \citep{euclid_2013} or DESI \citep{desi_2015}.
The accurate description of the velocity and density field around voids is
critical to properly perform cosmological test like the Alcock-Paczynski
\citep{lavaux_voids_2010,sutter_first_2012} or the Integrated Sachs-Wolfe
effect \citep{granett_isw_2008,papai_isw_2011,hernandez-monteagudo_signature_2013,
ilic_detection_2013,cai_cmb_2014}, and our results could become relevant on
this context contributing to the field by including a possible scenario for
large--scale flows based on the concordance cosmological model.

The non--negligible and coherent mutual void bulk motions observed here could
contribute to the motion of clusters and galaxies at intermediate densities, in
filaments or walls.
We stress the fact that bulk velocities of voids in observational data are in
concordance with the theoretical results.  We also note that the linear
approximation used to derive velocities are suitable for the analysis of the
bulk motion of voids in the observations (see Appendix \ref{appA}).


\section*{Acknowledgments}

We thank helpful comments and suggestions from the anonymous Referee, which
have improved and clarified this work.
This work has been partially supported by Consejo de Investigaciones
Cient\'{\i}ficas y T\'ecnicas de la Rep\'ublica Argentina (CONICET) and the
Secretar\'{\i}a de Ciencia y T\'ecnica de la Universidad Nacional de C\'ordoba
(SeCyT). 
Funding for the SDSS and SDSS-II has been provided by the Alfred P. Sloan
Foundation, the Participating Institutions, the National Science Foundation,
the U.S. Department of Energy, the National Aeronautics and Space
Administration, the Japanese Monbukagakusho, the Max Planck Society, and the
Higher Education Funding Council for England. The SDSS Web Site is
http://www.sdss.org/.
The SDSS is managed by the Astrophysical Research Consortium for the
Participating Institutions. The Participating Institutions are the American
Museum of Natural History, Astrophysical Institute Potsdam, University of
Basel, University of Cambridge, Case Western Reserve University, University of
Chicago, Drexel University, Fermilab, the Institute for Advanced Study, the
Japan Participation Group, Johns Hopkins University, the Joint Institute for
Nuclear Astrophysics, the Kavli Institute for Particle Astrophysics and
Cosmology, the Korean Scientist Group, the Chinese Academy of Sciences
(LAMOST), Los Alamos National Laboratory, the Max-Planck-Institute for
Astronomy (MPIA), the Max-Planck-Institute for Astrophysics (MPA), New Mexico
State University, Ohio State University, University of Pittsburgh, University
of Portsmouth, Princeton University, the United States Naval Observatory, and
the University of Washington.
We thank Dr. Mario A. Sgr\'o for kindly providing the numerical
simulation and halo catalogue. 
Some of the plots presented in this work were made by using {\sc r} software
and post processed using {\sc inkscape}.
This research has made use of NASA's
Astrophysics Data System.


\bibliographystyle{mnras}
\bibliography{references}


\appendix
\section{The linearized peculiar velocity field} 
\label{appA}

The analysis of large--scale dynamics is limited by the lack of sufficiently
large samples of galaxies with measured peculiar velocities with reasonable
accuracy which requires the determination of distances independent of redshift.
   
For this reason, in order to study the motions of observational voids, we use a
reconstructed velocity field using linear theory by
\citet{wang_reconstructing_2009,wang_reconstructing_2012} in the SDSS volume. 
We have already adopted these linearized velocities to  estimate the bulk
motions of voids identified in the SDDS in \citetalias{dgl_coherencia_2016}
where we claim a strong alignment and small moduli difference between
linearized and fully non-linear velocities.  We provide here the details of the
analysis of the effects of using linearized velocities to estimate bulk void
velocities.

\begin{figure}
\centering
\includegraphics[width=0.45\textwidth]{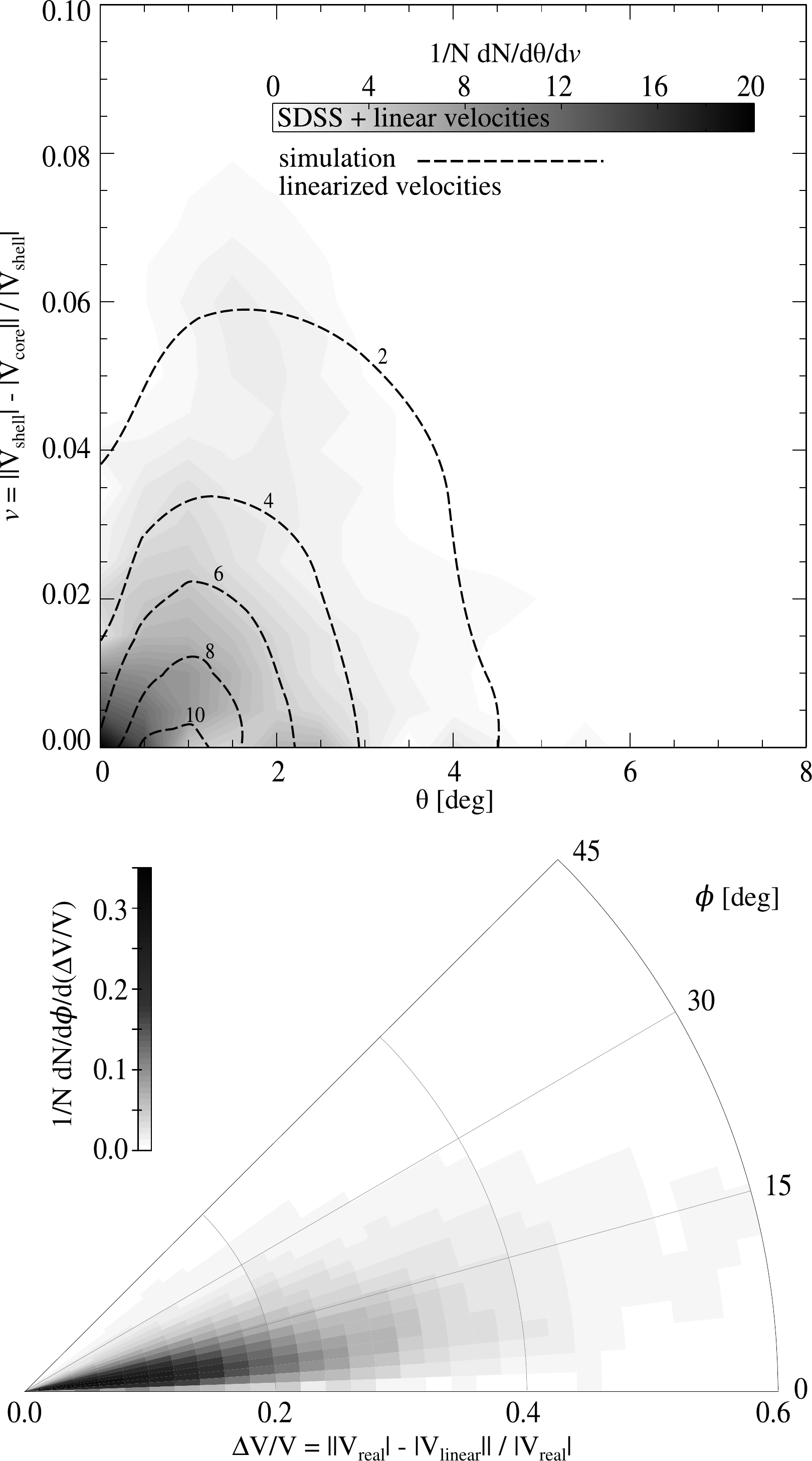}
\caption{
{\it Upper panel:} Probability density as a function of the angle between the
core and shell velocities ($\theta$) and the relative difference between both
velocities obtained from the SDSS+linearized velocity field.
The dashed lines corresponds to the same quantities computed through the
linearized velocities of the simulation. 
{\it Lower panel:} Polar diagram of the probability density  as a function of
the angle $\phi$ and the relative difference between the full and the
linearized velocity estimates of voids.
}
\label{fig:core} 
\end{figure}
  
We also showed in \citetalias{dgl_coherencia_2016} that the shell mean velocity
is representative of the velocity of the void core by comparing the velocity
obtained from simulated haloes at void walls ($0.8 < r/R_{\rm void} < 1.2$) to
the obtained using the mass distribution (particles) in the region $r/R_{\rm
void} < 0.8$. 
Here we show that this result is also valid for observational voids by
comparing their core and shell velocities.
In the upper panel of Fig. \ref{fig:core} we show an estimate of the density
of voids as a function of the angle between the core and the shell velocities
($\theta$) and the difference of the moduli between both velocities obtained
from the SDSS+linearized velocity field (shaded contours).
In dashed lines we show the same quantity computed for the linearized
velocities in the numerical simulation. 
These velocities where obtained applying the same linear reconstruction
procedure to the simulated halo catalogue. 
As it can be seen, the difference between the velocity directions are smaller
than 5 degrees, meanwhile the difference in magnitude is lower than 8 per cent.
To quantify the impact of this approximation to the true velocities of void
shells, we have implemented the method introduced by
\citet{wang_reconstructing_2009} to the density field traced by the massive
dark matter haloes in the simulation. 
Then, we can infer the linear velocity field in the simulation box in a
similar fashion than in the observational data.
The lower panel of Fig. \ref{fig:core} shows the probability density of voids
in the simulation as a function of the angle $\phi$ and the relative difference
between both velocities.
By inspection to this figure we can notice that the lineal reconstruction of
the velocity field has uncertainties in $\phi$ typically lower than 15 degrees
and 40 per cent in moduli.
Taking into account these results, we argue that linear theory provides a
suitable approximation to the actual void shell velocities, and therefore can
be used to study the motions of observational voids.

\end{document}